\documentclass[aps, prl, superscriptaddress, reprint]{revtex4-2}

\usepackage[utf8]{inputenc}
\usepackage[T1]{fontenc}
\usepackage{amsmath}
\usepackage{amsthm}
\usepackage{amssymb}
\usepackage{hyperref}
\usepackage{graphicx}
\usepackage{enumerate}
\usepackage{bm}
\usepackage{booktabs}
\usepackage[caption=false]{subfig}

\hypersetup{colorlinks=true, linkcolor=blue, citecolor=blue, urlcolor=blue}

\newcommand{\ev}{\text{ev}}

\newcommand{\keV}{\text{ keV}}

\newcommand{\eff}{\text{eff}}
\newcommand{\Br}{\text{Br}}
\newcommand{\tail}{\text{tail}}

\newcommand{\phys}{\text{phys}}

\newcommand{\He}{\text{He}}

\newcommand{\HeFour}{^{4}\He}

\newcommand{\HeFourIon}{\HeFour^{2+}}

\newcommand{\HIon}{\text{H}^+}
\newcommand{\DIon}{\text{D}^+}
\newcommand{\TIon}{\text{T}^+}
\newcommand{\Boron}{^{11}\text{B}}
\newcommand{\BIon}{\Boron^{5+}}

\begin{document}

% Date/Title
\title{Electrons quench ion tails during evaporative cooling in hot ion mode plasmas}
\date{\today}

% Author information
\author{Eric Palmerduca}
\email{ep11@princeton.edu}
\affiliation{Department of Astrophysical Sciences, Princeton University, Princeton, NJ 08540, USA}

\author{Hong Qin}
\affiliation{Department of Astrophysical Sciences, Princeton University, Princeton, NJ 08540, USA}
\affiliation{Department of Nuclear Engineering and Engineering Physics, University of Wisconsin-Madison, Madison, WI 53706, USA}

\author{Nathaniel J. Fisch}
\affiliation{Department of Astrophysical Sciences, Princeton University, Princeton, NJ 08540, USA}

\begin{abstract}
It is shown how evaporative cooling in hot plasma is essentially different from evaporative cooling in other media, such as neutral gases or Bose-Einstein condensates. The fundamental difference in plasmas arises both from the large mass ratio between electrons and ions in fully ionized plasma and the unusually sensitive dependence of plasma collisionality on speed. 
Thus, a hot ion mode plasma ($T_e < T_i$) is shown to support a distinctive evaporative cooling regime, where a strong reduction in the ion evaporation rate appears. This new regime may have application to approaches to economical nuclear fusion, where the ion tail  plays an outsized role in the fusion reaction rate.
\end{abstract}

\maketitle

\emph{Introduction}: Evaporative cooling is a kinetic process in which one process accomplishes the preferential removal of particles from the high-energy tail. To sustain the evaporation, a second process, typically collisional diffusion, continually repopulates the depleted tail of the distribution, thereby reducing the mean energy of the bulk distribution.
This principle appears in familiar natural processes, such as the cooling of liquids through surface evaporation. 
It has been exploited particularly effectively in atomic physics \cite{ketterle1996a, luiten1996}. 
Forced evaporation  enables neutral gases to reach ultracold temperatures below $1 \text{ nK}$ \cite{leanhardt2003}.
It is a critical step in the standard method of producing Bose-Einstein condensates \cite{Anderson1995}. 
The same kinetic process has been used in nonneutral plasmas, particularly by the ALPHA collaboration to evaporatively cool antiprotons and positrons to extremely cold temperatures for the synthesis of antihydrogen \cite{andresen2010,andresenThesis,Danielson2015,Ahmadi2017,Ahmadi2018}. 
Evaporative cooling has also been studied at higher temperatures in nonneutral trapped-ion systems, such as electron beam ion sources and traps \cite{marrs1999,kinugawa2001} and Penning traps \cite{hobein2011}.

In all these settings, a single population  determines both the equilibrium distribution and the collisional supply of particles to the evaporating tail. 
A hot, quasineutral plasma is fundamentally different. 
We show that the evaporation rate of one species can be heavily modified by the presence of a second non-evaporating species.

Two properties of plasma deliver its unique properties. 
First, the large ion-electron mass ratio allows the ion and electron bulks to remain nearly thermal but at substantially different temperatures, since \emph{thermal} particles have much faster intraspecies relaxation compared to interspecies relaxation. 
The large mass ratio  creates an energy interval, encompassing essentially the entire ion tail, in which ions are fast compared with thermal ions but slow compared with thermal electrons: $v_{th,i} \ll v \ll v_{th,e}$. 
Second, the Coulomb collision frequency in plasma is highly sensitive to the tail velocities. 
In the ion tail, the ion-ion energy diffusion coefficient falls as $D_{i} \sim E^{-1/2}$ (since $v_{th,i}\ll v$) while the ion-electron diffusion grows as $D_e \sim E$ (since $ v \ll v_{th,e}$). Thus, while ion-ion collisions control the ion bulk, ion-electron collisions can become important or even dominate the ion tail dynamics. 

This control is  particularly apparent in the hot ion mode, in which $T_e < T_i$. 
The hot ion mode is important in Deuterium-Tritium (DT) magnetic confinement fusion, where the top experimental results to date have all been obtained with this mode supported by auxiliary heating \cite{bell1995, keilhacker1999, maslov2023, vianello2026}. In a DT fusion reactor, achieving the hot ion mode requires channeling energy from the alpha particles \cite{fisch1992,fisch1994,fisch2006alpha}. 
For aneutronic fusion, such as proton-boron 11 (pB11) fusion, where the reactivity is already small, the hot ion mode is absolutely essential \cite{hay2015,putvinski2019,ochs2022,Liu2025}, not only because of the greater fusion reactivity for the same total confined pressure, but also because of the reduced radiative losses.
While a hot ion mode  is advantageous for DT fusion, for pB11 fusion it is critical, with superthermal protons disproportionately affecting fusion reactivity  \cite{kolmes2022}. In the hot ion mode regime, with larger ion energy, tail ions become increasingly affected  by collisions with the colder electron bath. Thus, evaporative cooling of ions, to the extent that reactant tail ions are affected, may be particularly relevant for pB11 fusion.

To expose the fundamental differences and new phenomena that emerge in evaporative cooling in plasma, as opposed to other media, particularly in the hot ion mode, we limit our focus to ions in a radial potential well.
Such potential wells can  occur in open magnetic field configurations.  
For example, under centrifugal confinement, but in an inverted mirror geometry, ions are confined isotropically up to the centrifugal potential \cite{Bekhtenev80}.  
The centrifugal potential could equivalently be replaced by a ponderomotive potential.  
Similarly, other barriers in open magnetic fields may confine ions below a maximum energy, such as electrostatic barriers \cite{rider1995}  or multiple mirror barriers \cite{Logan72,Miller2023}, with similar evaporative cooling effects.

\emph{Cooling by ion evaporation in plasma}:
To isolate the collisional mechanism, we consider ion evaporation in a weakly coupled, two-temperature ion-electron plasma in which ions with energy above a cutoff $E_0$ are removed at a finite rate. This one-dimensional reduction retains the competition between ion-ion relaxation, ion-electron relaxation, and evaporative loss while making the kinetic problem amenable to asymptotic analysis. At large cutoff energy, the solution separates the electron-modified distribution supplying the tail from the additional depletion caused by evaporation. This separation yields an explicit power-loss formula whose factors identify the competing physical effects and the conditions under which electron quenching controls evaporation.

Evaporated particles and energy are resupplied to the bulk $E \ll E_0$, modeling a steady-state ion distribution with finite evaporative loss rate. We identify an electron-quenched regime in which the cooler electron bulk suppresses the ion tail at energies well below those for which ion-electron collisions dominate diffusion. In a case of proton evaporation using pB11 fusion parameters, this effect can suppress the evaporated power by an order of magnitude. Fokker-Planck (FP) simulations reproduce the predicted distribution and evaporated power.

Consider a homogeneous, isotropic, weakly coupled, quasineutral plasma with fixed Maxwellian electrons with temperature $T_e$ and a single ion species of mass $m_i$, charge $Ze$, density $n_i = n_e/Z$, and temperature $T_i$. We generally assume $T_i > T_e$, modeling the hot ion mode, but the analytical treatment is valid for $T_i < T_e$. The ion energy distribution function $F(E,t)$ is normalized such that $\int_0^\infty F\, dE = n_i$. Particles above $E_0 = \eta_0 T_i$ are removed at a single particle rate $\gamma_{\ev}$, modeling evaporation. The energy space FP equation can be expressed as
\begin{gather}
    \frac{\partial F}{\partial t} = - \frac{\partial J}{\partial E} - \gamma_{\ev} \Theta(E - E_0)F + S(E) \label{eq:main_FP}\\
    J = J_i + J_e\\
    J_s = - D_s(E)\left[\frac{\partial F}{\partial E} + \left( \frac{1}{T_s}  - \frac{1}{2E}\right)F\right].
    \label{eq:J_s}
\end{gather}
Here, $D_s$ and $J_s$ are the energy-space ion diffusion coefficient and flux due to ion-$s$ collisions, and $\Theta$ is the step function. $S$ is a source which resupplies particles and energy lost via evaporation to the ion bulk, representing those sources which maintain a steady-state hot ion mode. The detailed form of $S$ is given in the End Matter and is used for FP simulation; for the analytic treatment it is only relevant to note that $S$ is nonzero only for energies $E \lesssim 3.5T_i < E_0$, below the cutoff for evaporation. For the analytic theory, we use a linear Landau test-particle operator for ion collisions against fixed Maxwellian electron and ion baths. In simulations, ion-ion collisions are more precisely treated using the nonlinear isotropic 1D Landau ion-ion collision operator. 

For $3T_i \lesssim E \ll (m_i/m_e)T_e$, the ratio of the diffusion coefficients is
\begin{gather}
    r_D(\eta) = \frac{D_e}{D_i} \simeq \left( \frac{\eta}{\eta_D} \right)^{3/2}, \qquad     \eta = \frac{E}{T_i} \\
    \eta_D = \left[\frac{3\sqrt{\pi}}{4}Z \frac{\ln \Lambda_{ii}}{\ln \Lambda_{ie}} \sqrt{\frac{m_iT_e}{m_eT_i}} \right]^{2/3} \label{eq:eta_D}
\end{gather}
$\eta_D$ is the normalized energy at which $D_e = D_i$. At this energy, electrons significantly enhance diffusion, and naively one might expect that the cutoff energy $\eta_0 = E_0/T_i$ must be greater than or equal to $\eta_D$ for electron modifications to be significant. However, we show that electron-induced modifications to the ion distribution can occur at much lower cutoff energies.

Our goal is to solve for the steady-state solution of Eq. (\ref{eq:main_FP}), denoted by $F(E)$, and the evaporated power $P_\ev = \int_{E_0}^{\infty} \gamma_{\ev}EF(E)\, dE$. In the limit of a large evaporative cutoff, $\eta_0 \gg 1$, evaporation should only modify the tail distribution with $\eta \gtrsim \eta_0$, and thus the deviation can be treated as a perturbation on the zero-evaporation ion distribution $\bar{F}$. It is essential to establish the correct zero-evaporation $\bar{F}$, as it can be significantly non-Maxwellian in the tail. Indeed, with no evaporation and considering energies $E > 3.5T_i$ such that $S(E) = 0$, the steady-state FP equation is $J[\bar{F}] = 0$, which has solution
\begin{gather}
    \bar{F}(E) = F_{M,i}(E)e^{-\bar{\chi}(E)} \label{eq:barF}\\
    \bar{T}_{\eff}(E) = \frac{D_i(E) + D_e(E)}{D_i(E)/T_i + D_e(E) / T_e} \\
    \bar{\chi}(E) = \int_{E_m}^E \left[\frac{1}{\bar{T}_{\eff}(E')} - \frac{1}{T_i}\right]dE'. \label{eq:chi_def}
\end{gather}
Here, $F_{M,i} = \sqrt{4E/\pi}n_i T_i^{-3/2}\exp(-E/T_i)$ is the Maxwellian ion distribution and $E_m = 4T_i$ is a matching energy chosen so it is above the range where the source $S$ applies but small enough that the distribution is still approximately Maxwellian. The matching condition is $\bar{F}(E_m) = F_{M,i}(E_m)$. The effective temperature of the zero-evaporation tail, $\bar{T}_\eff$, is essentially $T_i$ in the thermal bulk. However, $D_e$ grows faster than $D_i$ for high energy, causing $\bar{T}_\eff$ to approach $T_e$ in the extremely high energy tail. However, even at lower energies where $D_e(E)/D_i(E) \ll 1$ and $\bar{T}_\eff$ is only slightly smaller than $T_i$, Eq. (\ref{eq:chi_def}) shows that this slight suppression can accumulate over a large energy range into large suppression of the ion tail. We denote by $\eta_\chi$ the $\eta$ at which $\bar{\chi}(\eta_\chi) = 1$, the energy at which the Maxwellian is suppressed by a factor of $e$. Table \ref{tab:eta-crossovers} shows how at target parameters for pB11 fusion in the hot ion mode, $\eta_\chi$ is substantially smaller than $\eta_D$, indicating that the ion tail is quenched by electrons before electrons substantially enhance ion diffusion. These values also reflect the general trend apparent from Eq. (\ref{eq:eta_D}), namely that electrons have a stronger influence on the tails of lower mass, lower $Z$ ions such as the hydrogen isotopes.

\begin{table}[htbp]
    \centering
    \caption{$\eta_\chi$ and $\eta_D$ for
    $T_e=150\,\mathrm{keV}$, $T_i=300\,\mathrm{keV}$, and
    $n_e=10^{20}\,\mathrm{m}^{-3}$ for various fully ionized species}
    \label{tab:eta-crossovers}
    \begin{tabular}{lccccc}
        \toprule
        & $\HIon$ & $\DIon$ & $\TIon$ & $\HeFourIon$ & $\BIon$ \\
        \midrule
        $\eta_{\chi}$ & 7.9 & 8.7 & 9.2 & 12.0 & 19.9 \\
        $\eta_D$      & 12.0 & 15.1 & 17.3 & 29.6 & 74.3 \\
        \bottomrule
    \end{tabular}
\end{table}

\emph{Asymptotic evaporation law:} Finite-rate evaporation can be treated as a modification to the zero-evaporation solution $\bar{F}$ in the evaporation energy region $E \sim E_0$, factoring the full solution as
\begin{equation}
    F(E) = \bar{F}(E)h(E).
\end{equation}
For $\eta_0 \gg 1$, freeze $D=D_i+D_e$ and $\bar{T}_\eff$ in the vicinity of $E_0$, where $|E - E_0| \sim O(T_i)$. With $D_0 = D(E_0)$, $\bar{T}_{\eff,0} = \bar{T}_\eff(E_0)$, and defining the normalized evaporation rate $\bar{\gamma}_0 = \gamma_\ev \bar{T}_{\eff,0}^2/D_0$, the solution of Eq. (\ref{eq:main_FP}) with $h$ and $J$ continuous is
\begin{gather}\label{eq:h}
    h(E) = 
    \begin{cases}
        1 - A_{\ev}(\bar{\gamma}_0)e^{(E-E_0)/\bar{T}_{\eff,0}}  & E<E_0 \\
        [1 - A_\ev(\bar{\gamma}_0)]e^{-(E-E_0)/T_{\ev}} & E \geq  E_0.
    \end{cases} \\
    T_{\ev} = \frac{2\bar{T}_{\eff,0}}{(\sqrt{1+4\bar{\gamma}_0} - 1)} \\
    A_\ev(\bar{\gamma}_0) = \frac{\sqrt{1+4\bar{\gamma}_0} -1}{\sqrt{1+4\bar{\gamma}_0} + 1}.
\end{gather}
$A_\ev = 1-h(E_0)$ is the evaporation induced depletion at the cutoff, ranging continuously from slow extraction $(A_\ev \rightarrow 0)$ to diffusion-limited loss $(A_\ev \rightarrow 1)$. $T_\ev$ is the evaporation induced suppression of the ion tail temperature, related to the full tail temperature by $T_{\tail} = (\bar{T}_\eff^{-1} + T_\ev^{-1})^{-1}$.

\begin{figure}[htbp]
	
    \centering
    \includegraphics[width=0.48\textwidth]{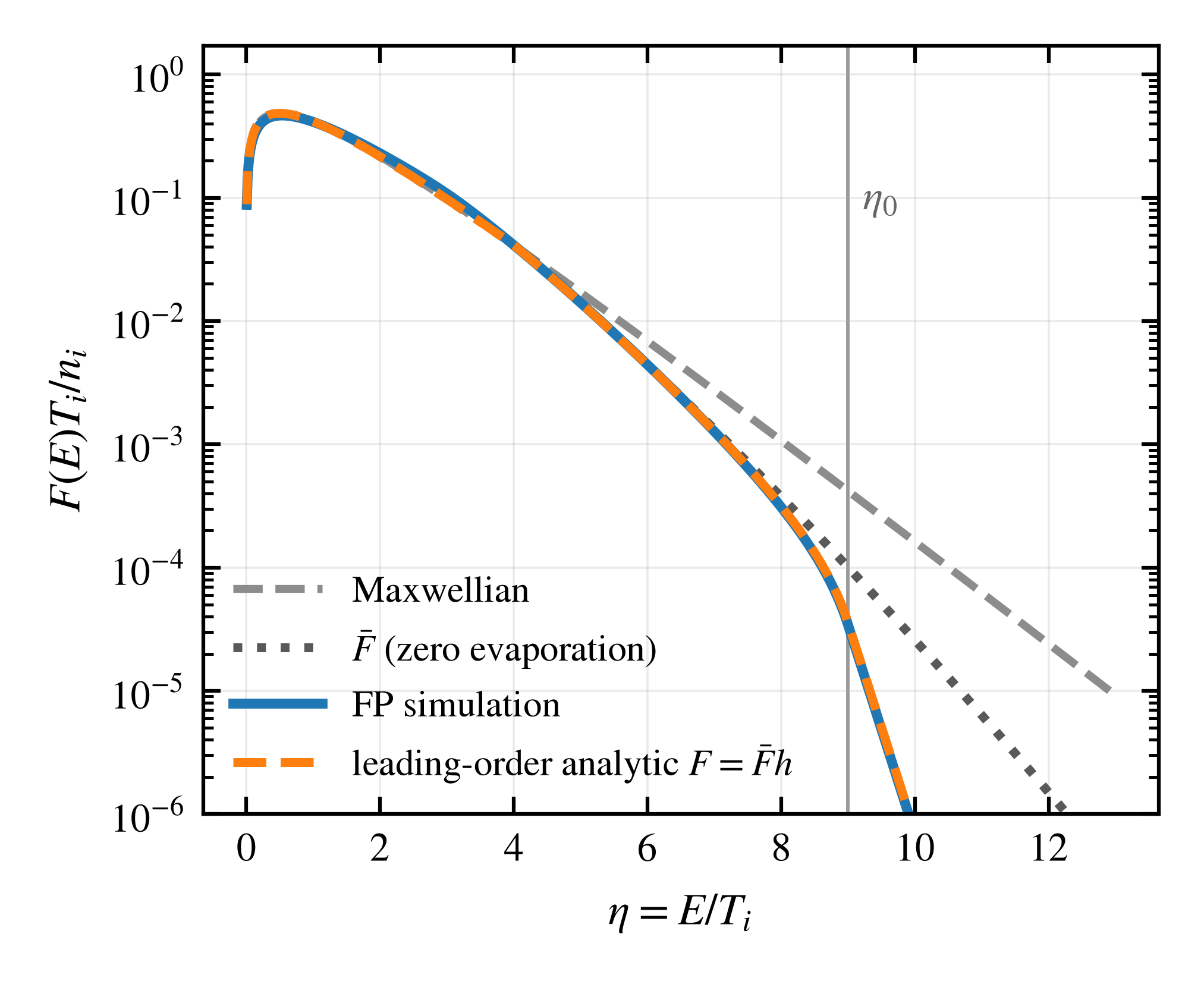}
	\caption{Analytic estimate and FP simulation of the normalized steady-state ion distribution $F(E)T_i/n_i$ during evaporative cooling of a proton-electron plasma with $T_i = 300 \keV$, $T_e = 150 \keV$, $n_e = 10^{20} \,\text{m}^{-3}$, $\gamma_0 \doteq \gamma_{\ev} T_i^2/D_0 = 10$, and $\eta_0 = 9$. A $300 \keV$ Maxwellian (gray dashed) and the quenched zero-evaporation distribution $\bar{F}$ (gray dotted) are shown for reference.}
	\label{fig:FP_proton_eta9}
\end{figure}

\begin{figure}[htbp]
	
    \centering
    \includegraphics[width=0.48\textwidth]{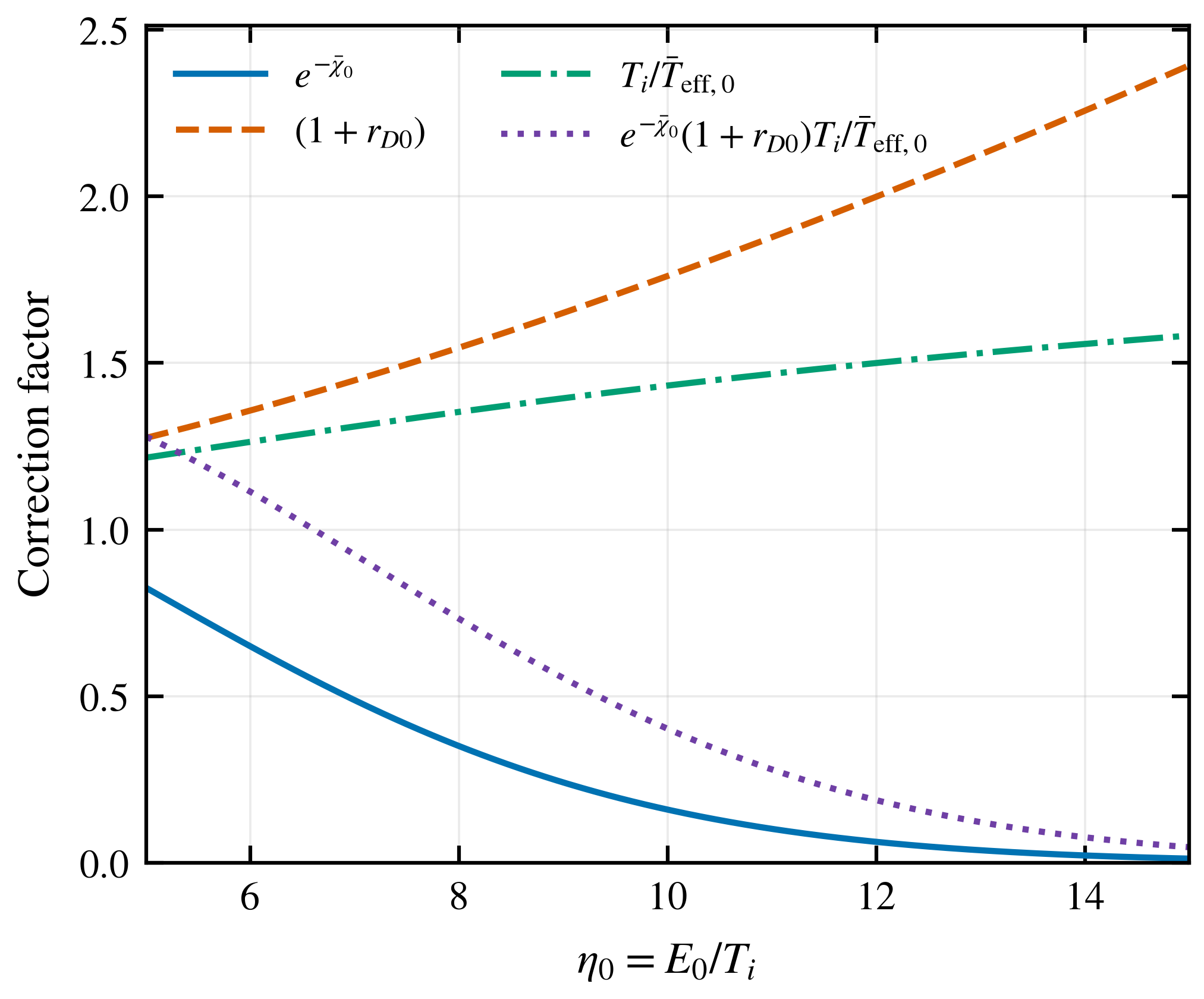}
    \label{fig:hot_ion_proton}

	\caption{$T_e < T_i$ modification factors to the evaporated ion energy flux from Eq. (\ref{eq:PevFactored}) in a proton-electron plasma. The evaporation cutoff $\eta_0 = E_0 / T_i$ is varied while $T_i = 300 \keV$, $T_e = 150 \keV$, and $n_e = 10^{20} \,\text{m}^{-3}$ are fixed.}
	\label{fig:hot_ion_mode_effects}
\end{figure}

\begin{figure*}[t]
\centering

\subfloat[][]{%
    \label{fig:P_ev_Comparison_vals}%
    \includegraphics[width=0.48\textwidth]
    {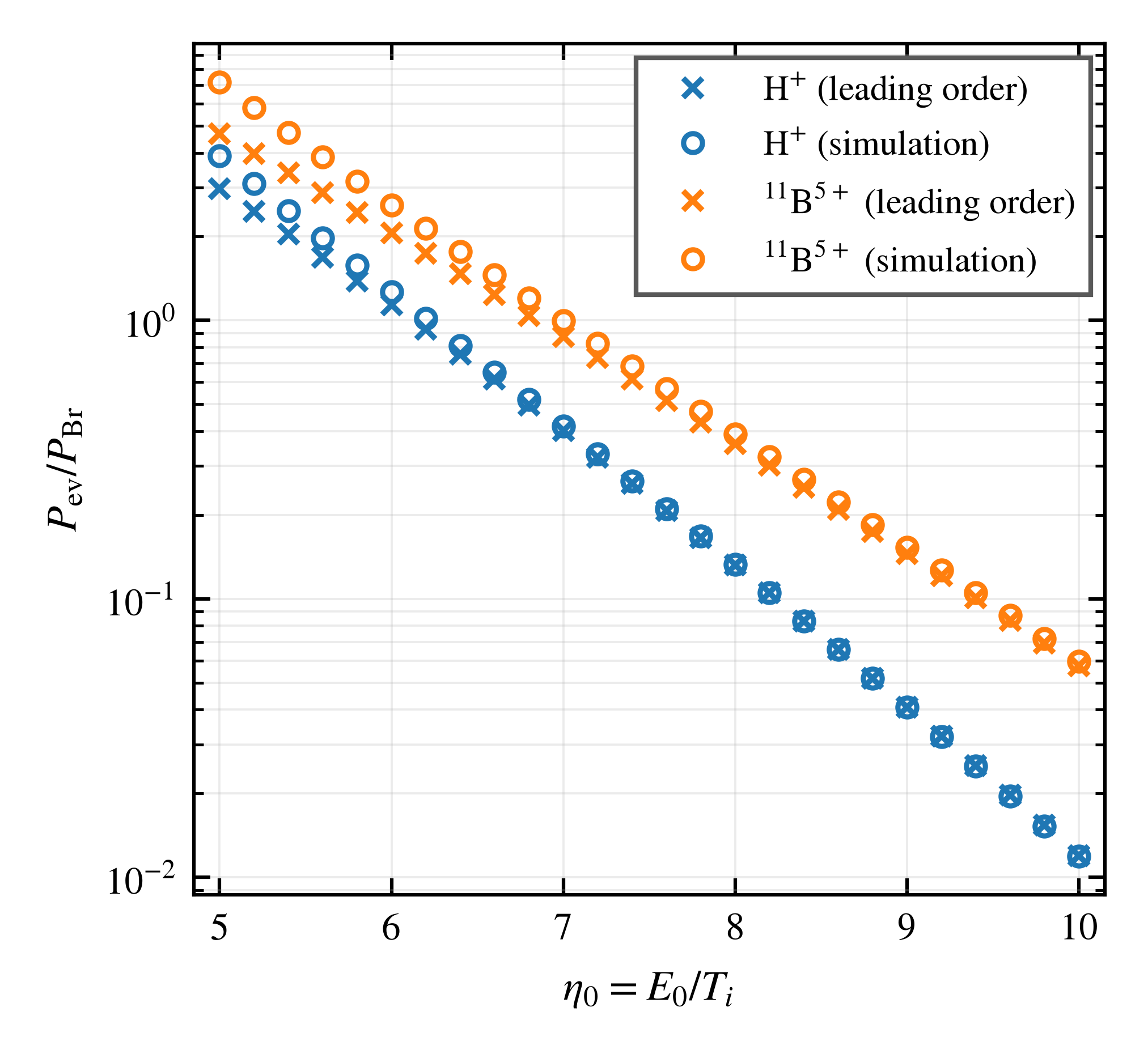}%
}
\hfill
\subfloat[][]{%
    \label{fig:P_ev_Comparison_error}%
    \includegraphics[width=0.48\textwidth]
    {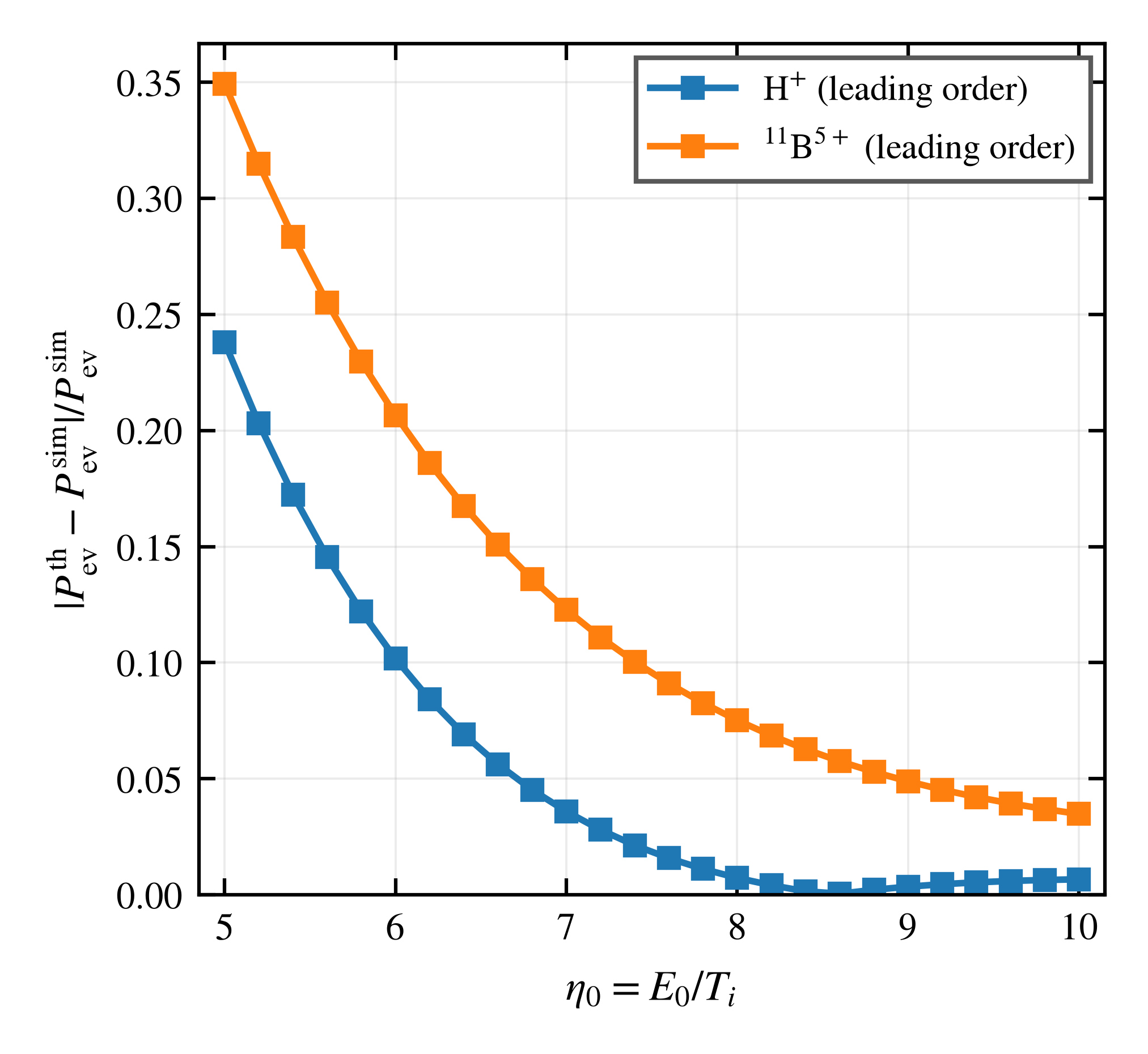}%
}

\caption{Comparison of the simulated and analytic estimates of the evaporated power as a function of the evaporation cutoff
$\eta_0 = E_0/T_i$. The physical parameters are
$T_i = 300 \keV$, $T_e = 150 \keV$, $n_e = 10^{20}\,\text{m}^{-3}$,
and normalized evaporation strength $\gamma_0 \doteq \gamma_\ev T_i^2 / D_0 = 10$. (a) Evaporated power in units of the electron bremsstrahlung and (b) absolute value of relative error vs.\ evaporation cutoff $\eta_0$.}
\label{fig:P_ev_Comparison}
\end{figure*}

The volumetric particle loss rate is $\Gamma_\ev = J(E_0)$ and the evaporated power density is $P_\ev = \gamma_{\ev} \int_{E_0}^\infty E F(E) \, dE$. The mean energy of an evaporated particle is $E_0 + T_\tail = E_0[1+O(\eta_0^{-1})]$, thus to leading order
\begin{align}
    P_{\ev} &= P_*\eta_0 e^{-\eta_0}A_{\ev}(\bar{\gamma}_0)\underbrace{e^{-\bar{\chi}_0}(1 + r_{D0})\frac{T_i}{\bar{T}_{\eff,0}}}_{\text{electron modification}} \label{eq:PevFactored}\\
    P_*   &= \frac{\sqrt{2}e^4 n_e^2 Z^2 \ln \Lambda_{ii}}{4\pi^{3/2}\epsilon_0^2\sqrt{m_iT_i}}
\end{align}
This form allows for easy identification of physical effects. If we consider the fast evaporation limit, $\bar{\gamma}_0 \rightarrow \infty$ and ignore ion-electron collisions, then the evaporated power is $P_* \eta_0 e^{-\eta_0}$. This limit is consistent with the evaporation rate previously found for nonneutral antiproton plasmas \cite{andresenThesis} in the absence of an externally applied potential. $A_\ev(\bar{\gamma}_0)$ measures the effect of finite evaporation, representing the fraction of ions at energy $E_0$ which have been evaporated. The bracketed terms show three modifications caused by ion-electron collisions. $(1+r_{D0})$ accounts for the electron enhancement of the diffusion coefficient. $e^{-\bar{\chi}_0}$ and $(T_i/ \bar{T}_{\eff,0})$ account for two opposing $T_e < T_i$ effects. $e^{-\bar{\chi}_0}$ describes the suppression of the ion tail population due to the cumulative effect of an effective temperature which decreases with $E$. On the other hand, the logarithmic slope of the distribution at the cutoff is steeper by a factor of $T_i / \bar{T}_{\eff, 0}$ compared with the $T_i = T_e$ case, enhancing the diffusive flux into the evaporation region. The diffusive enhancement effects grow algebraically in $\eta_0$ while the effect of the ion tail suppression $e^{-\bar{\chi}_0}$ grows exponentially and thus the latter becomes the dominant effect of electrons on ion evaporation. This explicit identification and separation of the physical mechanisms underlying evaporation is an advantage of the analytic treatment, providing physical insight that is not directly available from numerical solutions alone.

Figures \ref{fig:FP_proton_eta9} and \ref{fig:hot_ion_mode_effects} illustrate why it is essential to include the electron quenching effect when $T_e < T_i$. Figure \ref{fig:FP_proton_eta9} shows the analytic and simulated steady-state ion distribution functions for a proton-electron plasma using target parameters for pB11 fusion ($T_i = 300 \keV$, $T_e = 150 \keV$, $n_e = 10^{20} \,\text{m}^{-3}$) with evaporation above $\eta_0 = 9$. The root mean square error is $4.3\, \%$ over the range $0 < E/T_i < 12$. For energies below $\eta_0$, the ion tail follows $\bar{F}$, the quenched zero-evaporation distribution, rather than the $300 \keV$ Maxwellian inferred from the ion bulk. The magnitudes of the various electron-induced correction factors are shown in Figure \ref{fig:hot_ion_mode_effects} as a function of $\eta_0$. At $\eta_0 = 6.9$, the ion tail quenching from $e^{-\bar{\chi}}$ reduces $P_\ev$ by a factor of $2$, while at $\eta_0 = 11$ it is reduced by a factor of $10$. Above $\eta_0 = 6.6$, the tail quenching effect outweighs the combined effects of the enhanced diffusion coefficient and the steeper distribution slope.

To characterize the absolute size of the evaporative loss, we normalize $P_\ev$ to the relativistic electron bremsstrahlung loss power $P_\Br$ in an optically thin plasma quoted in Ref. \cite{munirov2023}. For advanced fuel fusion, $P_\Br$ sets the characteristic power loss rate, and thus $P_\ev / P_\Br$ serves as a figure of merit to indicate the relative effect of evaporative losses on the overall power balance. Since both $P_\ev$ and $P_\Br$ are proportional to $n_e^2$, their ratio is insensitive to density apart from a very weak dependence on the Coulomb logarithms in $P_\ev$.

We tested the asymptotic theory against a one-dimensional energy-space FP solver using the full Landau ion-ion collision operator, a fixed Maxwellian electron bath, and a particle and energy maintaining source $S$ in the ion bulk, designed to approximately preserve the Maxwellian profile in the bulk; see the End Matter for precise details. Figure \ref{fig:P_ev_Comparison} shows increasingly good agreement in the $\eta_0 \gg 1$ limit, with an error below $5\%$ for both protons and boron 11 ions at $\eta_0 = 10$.

\emph{Discussion and Conclusion}:
We showed how evaporative cooling in hot plasma is essentially different from evaporative cooling in other media. The large mass difference between ions and electrons allows for evaporative cooling regimes which strongly depend on multiple species, a situation which does not arise in the standard settings for evaporative cooling. In particular, the evaporative ion cooling regime in a hot ion mode plasma was shown to be fundamentally affected by the cooler electrons, leading to sharp depletion in the hot ion tail. The isotropic model used admits an asymptotic treatment that explicitly identifies and separates the effects of finite-rate evaporation, electron-enhanced diffusion, steepening of the distribution at the cutoff, and cumulative depletion of the ion tail. This analytic separation explains why the exponential suppression of the tail dominates the competing algebraic enhancements of the evaporative flux, providing physical insight beyond the combined effect observed from simulations.

Although the examples emphasize hot ions, the equations presented in this letter apply equally well in the case of $T_e > T_i$. In that case, $\bar{\chi} < 0$ and the electron bath enhances the ion tail and the evaporated power. We also emphasize that we have included finite evaporation effects by allowing $\gamma_\ev$ to be finite rather than assuming particles above $E_0$ are instantly removed as is standard in the evaporative cooling literature \cite{ketterle1996a, luiten1996, marrs1999, andresen2010,andresenThesis}. This is typically unnecessary to model neutral gas and nonneutral plasma traps, however, finite evaporation has been identified as a currently-missing correction needed to accurately model evaporative cooling in some strongly coupled ultracold neutral plasmas \cite{witte2017}. While we only provide the finite-evaporation theory in the weakly coupled case, the general approach likely extends to the strongly coupled case as well. 

The essential result is that while electrons have little effect on thermal ions, they can have a substantial effect on the ion tail, effectively quenching the tail in the hot ion mode. Theory and simulations predict that this quenching effect limits the active or passive evaporative cooling rate in hot ion mode plasmas. For evaporation of sufficiently hot ions, this suppression dominates the modest diffusive enhancement due to ion-electron collisions. Evaporative cooling remains comparatively unexplored in quasineutral plasmas, and the present results provide a concrete example of how multicomponent kinetics can give rise to new evaporation regimes absent from traditional one-component treatments.

\begin{acknowledgments}
    \emph{Acknowledgments}---This work was supported by the Eric and Wendy Schmidt Transformative Technology Fund and DOE DE-SC0016072.
\end{acknowledgments}

\bibliography{evaporative_cooling}

\section*{End Matter}
\emph{Fokker--Planck simulations}: We solve Eq.~(\ref{eq:main_FP}) on a uniform energy grid using a conservative Chang-Cooper discretization. The nonlinear isotropic Landau ion-ion operator is evaluated from Rosenbluth potentials, with its coefficients recomputed from $F_i$ at every time step. Ion-electron collisions are treated with a fixed Maxwellian electron bath, so those coefficients are computed once. The ions are initialized with $F_{M,i}$, and the collisional and control fluxes vanish at $E=0$ and $E=E_{\max}$. We use
\begin{gather}
    \Delta E=T_i/80,\qquad \Delta t=0.01\tau_{ii},\\
    E_{\max}=\max(14T_i,E_0+4T_i),
\end{gather}
where the ion-ion collision time is
\begin{equation}
    \tau_{ii}=
    \frac{12\pi^{3/2}\epsilon_0^2\sqrt{m_i}T_i^{3/2}}
    {n_iZ^4e^4\ln\Lambda_{ii}},
    \label{eq:end_tau_ii}
\end{equation}
with $T_i$ in units of energy \cite{NRL_Formulary}. The solution is evolved to $t=7\tau_{ii}$, by which time it has reached steady state. The numerically determined power loss density is $P_\ev=\gamma_\ev\int_{E_0}^{E_{\max}}EF_i\,dE$.

\emph{Particle and energy source}: The simulations seek a steady state at fixed ion density and ion energy while retaining only the ion--ion collisions, ion--electron collisions, and evaporation as explicit physical processes. We thus include a source $S$ which accounts for an external supply of particles and energy which preserves the total particle number and energy of the plasma. We thus separate out the explicit physical processes in the FP equation:
\begin{align}
	\frac{\partial F_i}{\partial t}
	&= R_{\phys}[F_i] + S, \label{eq:appendix_fp_split} \\
	R_{\phys}[F_i]
	&\doteq C_{ii}[F_i]+C_{ie}[F_i,F_{M,e}]
	-\gamma_{\ev}\Theta(E-E_0)F_i
\end{align}
and define the particle and energy loss moments
\begin{gather}
    \dot n_\phys=\int_0^\infty R_\phys\,dE, \\
    \dot U_\phys=\int_0^\infty ER_\phys\,dE.
\end{gather}
Ion-ion collisions contribute to neither moment, ion-electron collisions contribute only to $\dot U_\phys$, and evaporation contributes to both. These losses are offset by
\begin{equation}
    S(E,t) = R_N(t)\psi(E)
    -\frac{\partial}{\partial E}\left[\alpha(t)EW(E)F_i(E,t)\right],
    \label{eq:end_source}
\end{equation}
where
\begin{gather}
    \psi(E) = \frac{F_M(E;n_{i0},T_{i0})W(E)}
    {\int_0^\infty F_M(E;n_{i0},T_{i0})W(E)\,dE}, \\
    W(E) = \frac{1}{2}\left[1-\tanh\left(\frac{E-3T_{i0}}
    {0.5T_{i0}}\right)\right]
\end{gather}
and $R_N(t)$ and $\alpha(t)$ are dynamically determined (see Eq. (\ref{eq:end_source_amplitudes})). The two terms in Eq. (\ref{eq:end_source}) describe particle and energy injection, respectively. The window function $W(E)$ confines the injection to the thermal bulk, cutting off injection above $\sim 3.5T_i$ so that there is no direct injection in the evaporation layer. The geometric motivation for these profiles is clearest without the window. For $W=1$ at the target state $(n_{i0},T_{i0})$, we have $\psi=F_M/n_i$. The Maxwellian energy distributions
\begin{equation}
    F_M(E;n_i,T_i) = \frac{2n_i\sqrt{E}}{\sqrt{\pi}T_i^{3/2}}
    e^{-E/T_i}
\end{equation}
form a two-dimensional manifold parameterized by $n_i$ and $T_i$, whose tangent vectors are
\begin{align}
   \frac{\partial F_M}{\partial n_i}&=\frac{F_M}{n_i},\\
    \frac{\partial F_M}{\partial T_i} &= -\frac{1}{T_i}\frac{\partial}{\partial E}(EF_M)
\end{align}
The unwindowed source can therefore be written as
\begin{equation}
    S=R_N\frac{\partial F_M}{\partial n_i}
    +\alpha T_i\frac{\partial F_M}{\partial T_i},
\end{equation}
so that over a short time step $\delta t$,
\begin{equation}
	F_M+S\delta t
	=F_M(E; n_i+R_N \delta t,
	T_i + \alpha T_i \delta t) + O(\delta t^2).
	\label{eq:appendix_tangent_step}
\end{equation}
Thus, to leading order, the source changes the density and temperature without modifying the bulk shape. The window removes this exact tangency to the Maxwellian manifold but confines the control to the bulk. With $\bar{E}_\psi=\int_0^\infty E\psi\,dE$ and $\mathcal A[F_i]=\int_0^\infty EW(E)F_i\,dE$, the vanishing boundary flux gives
\begin{equation}
    \int_0^\infty S \, dE=R_N,\qquad
    \int_0^\infty ES \, dE=R_N \bar{E}_\psi + \alpha\mathcal A[F_i].
\end{equation}
Consequently, setting
\begin{equation}
    R_N=-\dot n_\phys,\qquad
    \alpha=\frac{-\dot U_\phys-R_N \bar{E}_\psi}{\mathcal A[F_i]}
    \label{eq:end_source_amplitudes}
\end{equation}
holds both moments fixed.

\emph{Detailed-balance form}: For a one-dimensional energy-space drift--diffusion operator, write
\begin{equation}
    \left.\frac{\partial F}{\partial t}\right|_s=-\frac{\partial J_s}{\partial E},
    \qquad J_s=A_sF-\frac{\partial}{\partial E}(D_sF).
\end{equation}
A Maxwellian bath at $T_s$ must have the stationary ion distribution $F_M(E;T_s)\propto \sqrt{E}\exp(-E/T_s)$. Detailed balance, $J_s[F_M]=0$, therefore fixes the drift:
\begin{align}
    A_s&=\frac{dD_s}{dE}+D_s\frac{d\ln F_M}{dE}
    \\&=\frac{dD_s}{dE}+D_s\left(\frac{1}{2E}-\frac{1}{T_s}\right).
\end{align}
Substitution gives
\begin{align}
    J_s&=-D_sF_M\frac{\partial}{\partial E}\left(\frac{F}{F_M}\right)\\
    &=-D_s\left[\frac{\partial F}{\partial E}
    +\left(\frac{1}{T_s}-\frac{1}{2E}\right)F\right],
\end{align}
which is the form used in Eq. (\ref{eq:J_s}).
\end{document}